\documentclass[11pt,fleqn]{article}

\usepackage{amsmath,amssymb,amsthm,enumerate}%cite,

\newcounter{tbn}

\newcounter{mcasenum}

\newtheorem{theorem}{Theorem}

\newtheorem*{proposition*}{Proposition}
{\theoremstyle{definition}

\newtheorem{example}{Example}

}

\begin{document}

\begin{center}{\LARGE\bf
Group classification via mapping between classes:\\ an example  of semilinear
reaction--diffusion equations\\[0.5ex] with exponential nonlinearity}

\vspace{4mm}
{\large\it Olena Vaneeva}

\vspace{2mm}
{\it Institute of Mathematics of NAS of Ukraine, 3 Tereshchenkivska Str., Kyiv-4, 01601 Ukraine}

\vspace{2mm}
{\it vaneeva@imath.kiev.ua}
\end{center}

{\vspace{5mm}\par\noindent\hspace*{8mm}\parbox{146mm}{\small The group classification of a class of 
semilinear reaction--diffusion equations with exponential nonlinearity
is carried out using the technique of mapping between classes, which was recently proposed in
[O.O. Vaneeva, R.O. Popovych and C. Sophocleous, {\it Acta~Appl.~Math.}, doi:10.1007/s10440-008-9280-9, arXiv:0708.3457].
}\par\vspace{2mm}}

\section{Introduction}
There exist relatively few equations describing natural phenomena among a great number of partial differential
equations (PDEs). This begs the question what mathematical properties differ
equations describing physical processes from other
possible ones? It appears that large majority of equations of mathematical physics has
nontrivial symmetry properties (see a number of examples e.g. in~\cite{FN}). It means that manifolds of their solutions are invariant with
respect to multi-parameter groups of continuous
transformations (Lie groups of transformations) with a number of parameters.
Therefore, the presence of nontrivial symmetry properties is one of
such distinctive features (and very important one)!

In some cases the requirement of invariance of equations under a group
enables us  to select these equations from a
wide set of other admissible ones.
For example, there is the only one system of
Poincar\'e-invariant partial differential equations of first order
for two real vectors $\mathbf
E(x_0,\mathbf{x})$ and $\mathbf B(x_0,\mathbf{x})$, and this is the system of
Maxwell equations~\cite{FN}. The problem arises to single out equations having high symmetry
properties from a given class of PDEs.
A solution of so-called {\it group classification problem} gives an exhaustive solution of this problem.

There exist two main approaches of solving group classification problems. The first one is more algebraic
and based on subgroup analysis of the equivalence group of
a class of differential equations under consideration (see~\cite{LZh1,LZh2,LZh3,LZh4} for details).

The second approach involves the investigation of compatibility and the direct integration
of determining equations implied by the infinitesimal invariance criterion~\cite{Ovsiannikov1982}.
Unfortunately it is efficient only for classes of a simple structure, e.g.,
which have a few arbitrary elements of one or two same arguments.
A number of results on group classification problems investigated within the framework of this approach are
collected in \cite{Ibragimov1994V1;2} and other books on the subject.

To solve more group classification problems different tools have been recently developed.
One of them is to carry out group classification using appropriate
mapping of a given class to a one having a simpler structure. See the theoretical background of this approach and
the first example of its implementation in~\cite{VPS_2009}.

In this paper we perform the group classification of the class of semilinear reaction--diffusion equations with exponential nonlinearity
\begin{equation}\label{eqRDfghExp_n=0}
f(x)u_t=(g(x)u_x)_x+h(x)e^{mu}
\end{equation}
in the framework of this approach.  Here \mbox{$f=f(x)$, $g=g(x)$} and \mbox{$h=h(x)$}  are arbitrary smooth functions of the variable $x$,
$fgh\neq0$,   $m$ is an arbitrary constant. The linear case is
excluded from consideration as well-investigated (i.e., we assume $m\neq0$).
\section{Equivalence transformations and mapping of class (1) to a simpler one}
It is essential for group classification problems to derive the transformations which
preserve differential structure of a class under consideration and transform only arbitrary elements.
Such transformations are called {\it equivalence} ones and form a group~\cite{Ovsiannikov1982}.

There exist several kinds of equivalence groups. The simplest one is given by usual equivalence groups
which consist of the nondegenerate point transformations of independent and dependent variables
as well as transformations of arbitrary
elements of a class.  Here transformations of independent and dependent variables do not depend on arbitrary
elements. If such dependence arises then the corresponding equivalence group is called {\it generalized}. If
 new arbitrary elements are expressed via old ones in some nonpoint, possibly nonlocal, way (e.g. new
arbitrary elements are determined via integrals of old ones) then
the equivalence transformations are called {\it extended} ones. The first examples of a
generalized equivalence group and of an extended equivalence group are presented in~\cite{Meleshko1994} and~\cite{mogran},
respectively. See a number of examples
of different equivalence groups and their role in solving complicated group classification problems,
e.g., in~\cite{VPS_2009,VJPS2007a,VJPS2007b}.

\begin{theorem} 
The generalized extended equivalence group~$\hat
G^{\sim}$ of class~\eqref{eqRDfghExp_n=0}  consists of the transformations
\[
\begin{array}{l}
\tilde t=\delta_1 t+\delta_2,\quad \tilde x=\varphi(x),\quad
\tilde u=\delta_3u+\psi(x), \\[1ex]
\tilde f=\dfrac{\delta_0\delta_1}{\varphi_x}f, \quad
\tilde g=\delta_0\varphi_x g,\quad
\tilde h=\dfrac{\delta_0\delta_3}{\varphi_x}e^{-\frac{m\psi(x)}{\delta_3}}h,
\quad \tilde m=\dfrac{m}{\delta_3},
\end{array}
\]
where  $\varphi(x)$  is an arbitrary smooth function, and $\psi(x)=\delta_4\int\frac{dx}{g(x)}+\delta_5$.
Here $\delta_j,$ $j=0,1,\dots,5,$ are arbitrary constants,
$\delta_0\delta_1\delta_3\not=0$.
\end{theorem} 

The above transformations with $\delta_4=0$ form the usual
equivalence group of class~\eqref{eqRDfghExp_n=0}.

The presence of the arbitrary function $\varphi(x)$ in the equivalence
transformations from  $\hat G^{\sim}$ allows us to simplify the group classification problem of
class~\eqref{eqRDfghExp_n=0} via reducing the number of arbitrary elements and making its more convenient for
mapping to another class.

Thus, the transformation from the equivalence group $\hat G^{\sim}$
\begin{equation}\label{gauge_f=g}
\tilde t={\rm sign }(f(x)g(x))\,t,\quad \tilde x=\int\left|\frac{f(x)}{g(x)}\right|^\frac12dx, \quad
\tilde u=m\,u,
\end{equation}
connects~\eqref{eqRDfghExp_n=0} with the class
$\tilde f(\tilde x) \tilde u_{\tilde t}=(\tilde f(\tilde x)\tilde u_{\tilde x})_{\tilde x}+\tilde h(\tilde x) e^{\tilde u},$
with the new arbitrary elements  $\tilde f(\tilde x)=\tilde g(\tilde x)={\rm sign}(g(x))|f(x)g(x)|^\frac12$, 
 $\tilde h(\tilde x)=m\left|\frac{g(x)}{f(x)}\right|^\frac12h(x)$, \mbox{$\tilde m=1$}.

Without loss of generality, we can restrict ourselves to
the study of the class
\begin{gather}\label{class_f=g}
f(x)u_{t}= (f(x)u_x)_x+h(x)e^u,
\end{gather}
since
all results on symmetries and exact solutions for this class can be extended to
class~\eqref{eqRDfghExp_n=0} with transformation~\eqref{gauge_f=g}.

It is easy to deduce the generalized extended equivalence group for class~\eqref{class_f=g}
from Theorem 1 by setting
$\tilde f=\tilde g$, $f=g$ and $\tilde m=m=1$.
The results are summarized in the following theorem.

\begin{theorem} The generalized extended equivalence group~$\hat
G^{\sim}_1$ of class~\eqref{class_f=g} is formed by the transformations
\[
\begin{array}{l}
\tilde t=\delta_1^{\,2} t+\delta_2,\quad \tilde x=\delta_1x+\delta_3,\quad
\tilde u=u+\psi(x), \\[1.5ex]
\tilde f=\delta_0\delta_1^{\,2}f, \quad
\tilde h=\delta_0e^{-\psi(x)}h,
\end{array}
\]
where $\psi(x)=\delta_4\int\frac{dx}{f(x)}+\delta_5$; $\delta_j,$ $j=0,1,\dots,5,$ are arbitrary constants,
$\delta_0\delta_1\not=0$.
\end{theorem}

The next step is to change the dependent variable in class~\eqref{class_f=g}:
\begin{equation}\label{gauge}
v(t,x)=u(t,x)+\omega(x),\quad\mbox{where}\quad\omega(x)=\ln|f(x)^{-1}h(x)|.
\end{equation}
As a result, we obtain the class
\begin{equation}\label{class_vFH}
v_t=v_{xx}+F(x)v_x+\varepsilon e^{\,v}+H(x),
\end{equation}
where $\varepsilon={\rm sign}(f(x)h(x))$ and the new arbitrary elements $F$ and $H$ are expressed via the formulas
\begin{equation}\label{eqfF}
F=f_xf^{-1},\quad H=-\omega_{xx}-\omega_xF.
\end{equation}

All results on Lie symmetries and exact solutions of class~\eqref{class_vFH}
can be extended to class~\eqref{class_f=g} by the inversion of transformation~\eqref{gauge}.
See the  theoretical background in~\cite{VPS_2009}.

\section{Lie symmetries}
In the previous section  the group
classification problem of class~\eqref{eqRDfghExp_n=0} has been reduced to
the similar but simpler problem for class~\eqref{class_vFH}.
In this section we investigate Lie symmetry properties of class~\eqref{class_vFH}.
Then the obtained results are used
to derive the group classification of class~\eqref{class_f=g} that is equivalent 
to class~\eqref{eqRDfghExp_n=0} with respect to transformation~\eqref{gauge_f=g} from $\hat G^{\sim}$.

The group classification problem for class~\eqref{class_vFH} is solved in the
framework of the classical approach~\cite{Ovsiannikov1982}. All
necessary objects (the equivalence group, the kernel and all
inequivalent extensions of maximal Lie invariance algebras) are
found.

The usual equivalence group~$G^{\sim}$ of class~\eqref{class_vFH} is formed by the transformations
\[\begin{array}{l}
\tilde t=\delta_1^{\,2} t+\delta_2,\quad \tilde x=\delta_1x+\delta_3, \quad
\tilde v=v-\ln \delta_1^{\,2},\\[1ex]
\tilde F=\delta_1^{\,-1}F, \quad
\tilde H=\delta_1^{\,-2}H,
\end{array}\]
where $\delta_j,$ $j=1,2,3,$  are arbitrary constants, $\delta_1\not=0$.

The generalized extended equivalence group of class~\eqref{class_vFH} degenerates to the usual one.

The \emph{kernel} of the maximal Lie invariance algebras of equations from class~\eqref{class_vFH}
coincides with the one-dimensional algebra $\langle\partial_t\rangle$. It means that any equation from
class~\eqref{class_vFH} is invariant with respect to translations by $t$.

All possible $G^\sim$-inequivalent \emph{cases of extension} of the maximal Lie invariance
 algebras in class~\eqref{class_vFH} are exhausted by ones
presented in Table~1.
\begin{center}\renewcommand{\arraystretch}{1.7}
\textbf{Table 1.} The group classification of class~\eqref{class_vFH}
\\[1ex]
\begin{tabular}{|c|c|c|c|l|}
\hline
N&$F(x)$&$H(x)$&\hfil Basis of $A^{\rm max}$ \\
\hline
%0&$\forall$&$\forall$&$\partial_t$\\
%\hline
1&$\alpha x^{-1}+\mu x$&$\beta x^{-2}+2\mu $&$\partial_t,\,e^{-2\mu t}(\partial_t-\mu x\partial_x+2\mu \partial_v)$\\
\hline
2&$\alpha x^{-1}$&$\beta x^{-2}$&$\partial_t,\,2t\partial_t+x\partial_x-2\partial_v$\\
\hline
3&$\mu x$&$\gamma$&$\partial_t,\,e^{-\mu t}\partial_x$\\
\hline

4&$\lambda$&$\gamma$&$\partial_t,\,\partial_x$\\
\hline
5&$\mu x$&$2\mu $&$\partial_t,\,e^{-\mu t}\partial_x,\,e^{-2\mu t}(\partial_t-\mu x\partial_x+2\mu \partial_v)$\\
\hline
6&$\lambda$&$0$&$\partial_t,\,\partial_x,\,2t\partial_t+(x-\lambda t)\partial_x-2\partial_v$\\

\hline
\end{tabular}
\\[1ex]
\parbox{150mm}{Here   $\lambda\in\{0,1\}\bmod G^{\sim}$, $\mu=\pm1\bmod G^{\sim}$; $\alpha, \beta, \gamma$ are arbitrary constants,
$\alpha^2+\beta^2\neq0$.
In case 3 $\gamma\neq2\mu$, in case 4 $\gamma\ne0$.}
\end{center}

Now we are able to derive the group classification of class~\eqref{class_f=g} using the results of Table~1.
To find the cases of extension of the maximal Lie invariance
 algebras in class~\eqref{class_f=g} we should, at first,  to solve ODEs~\eqref{eqfF}
for each pair of functions $F$ and $H$ from Table~1. In such a way we will obtain the functions $f$ and $\omega$.
Then all corresponding  $h$ can be easily found from
the formula \[h(x)=\delta f(x) e^{\,\omega(x)}, \quad \delta=\pm1.\]

In Table~2 we list the general solutions of~\eqref{eqfF} which are connected with six pairs of
functions $F$ and $H$ presented by cases~1--6
of Table~1.
\begin{center}\renewcommand{\arraystretch}{1.7}
\textbf{Table 2.} The general solutions of equations~\eqref{eqfF}\\[1ex]
\begin{tabular}{|l|c|c|}
\hline
N&$f(x)$&$\omega(x)$\\
\hline
1&$c_0 x^{\alpha}
 e^{\frac{\mu}2x^2}$&$\int\bigl(c_1-\int(\beta x^{-2}+2\mu)x^{\alpha}
e^{\frac{\mu}2x^2}dx\bigr) x^{-\alpha} e^{-\frac{\mu}2x^2}dx+c_2$\\
\hline
$2|_{\alpha\neq1}$&$c_0 x^\alpha$&$\frac \beta{1-\alpha}\ln x+c_1x^{1-\alpha}+c_2$\\
\hline
$2|_{\alpha=1}$&$c_0 x$&$-\frac {\beta}2\ln^2x+c_1\ln x+c_2$\\
\hline
3&$c_0 e^{\frac{\mu}2x^2}$&$\int\bigl(c_1-\gamma\int e^{\frac{\mu}2x^2} dx\bigr)e^{-\frac{\mu}2x^2}dx+c_2$\\
\hline
$4|_{\lambda=1}$&$c_0 e^{x}$&$-\gamma x+c_1e^{-x}+c_2$\\
\hline
$4|_{\lambda=0}$&$c_0$&$-\frac {\gamma}2 x^2+c_1x+c_2$\\
\hline
5&$c_0 e^{\frac{\mu}2x^2}$&$\int\bigl(c_1-2\mu\int e^{\frac{\mu}2x^2} dx\bigr)e^{-\frac{\mu}2x^2}dx+c_2$\\
\hline
$6|_{\lambda=1}$&$c_0 e^{x}$&$c_1e^{-x}+c_2$\\
\hline
$6|_{\lambda=0}$&$c_0$&$c_1x+c_2$\\
\hline
\end{tabular}
\\[1ex]
\parbox{150mm}{Note that $\int e^{\frac{\mu}2x^2} dx=\frac{\sqrt{\pi}}{\sqrt{-2\mu}}\,{\rm Erf}\!\left(\frac12\sqrt{-2\mu}x\right)$,
where ${\rm Erf}(z)$ is the error function.  

$c_i$, $i=0,1,2,$ are arbitrary constants, $c_0\neq0$.}
\end{center}

Transformation~\eqref{gauge} is not a bijection since the preimage set of each equation from class~\eqref{class_vFH}
is a two-parametric family of equations from class~\eqref{class_f=g}.
Every such family consists of equations which are
equivalent with respect to the group $\hat G^\sim_1$ from Theorem 2 (see the proof in~\cite{VPS_2009}).
A classification list for class~\eqref{class_f=g} can be obtained from
a classification list for class~\eqref{class_vFH} by means of taking a single preimage for each
element of the latter list with respect to the mapping
realized by transformation~\eqref{gauge}. It means that we should choose partial solutions of equations~\eqref{eqfF}
from the general ones presented in Table~2 in order to obtain the group classification of class~\eqref{class_f=g} up to $\hat G^{\sim}_1$-equivalence.

\begin{example} 
The equation $v_t=v_{xx}+v_x+e^v+\gamma$ from class~\eqref{class_vFH} is the
image of the family of equations from
class~\eqref{class_f=g}
\begin{gather}\label{case3_eq1}
e^xu_t=(e^xu_x)_x+e^{-\gamma x+c_1e^{-x}+c_2} e^u
\end{gather}
with respect to the transformation $v=u-(1+\gamma)x+c_1e^{-x}+c_2$.

The simplest representative of this family
is the equation
\begin{equation}\label{case3_eq2}
e^{\tilde x}{\tilde u}_{\tilde t}=(e^{\tilde x}{\tilde u}_{\tilde x})_{\tilde x}+e^{-\gamma\tilde x}e^{\tilde u}.
\end{equation}
Theorem~2 implies 
that equations~\eqref{case3_eq1} and~\eqref{case3_eq2} are equivalent with respect to the transformation
$\tilde t=t$, $\tilde x=x$, $\tilde u =u+c_1e^{-x}+c_2$ from~$\hat G^{\sim}_1$.
Hence, knowing the maximal Lie invariance algebra or exact solutions of~\eqref{case3_eq2},
one can derive the basis elements of the maximal Lie invariance algebra and exact solutions of equation~\eqref{case3_eq1}
that has more complicated coefficients.
\end{example} 

Therefore, to complete the group classification of 
class~\eqref{class_f=g} with respect to its equivalence
group~$\hat G^{\sim}_1$,  we should set, e.g., $c_1=c_2=0,$ $c_0=1$ 
in the functions $f$ and $h$
and construct the basis operators of the maximal Lie 
invariance algebras for equations from~\eqref{class_f=g} 
with such $f$ and $h$
using the formula
\[X=\tau\partial_t+\xi\partial_x+(\eta-\xi\omega_x)\partial_u.\]
Here $\tau,$ $\xi$ and $\eta$ are coefficients 
of  $\partial_t,$
$\partial_x$ and $\partial_v$ in infinitesimal generators 
from Table~1. $\omega_x=\frac{d\omega}{dx},$ where the corresponding values of 
$\omega$ connected with $f$ and $h$ via~\eqref{gauge} are  listed in Table~2  .

The obtained results are collected in Table~3. The first number
of each case indicates the associated case of Table~1.
\begin{center}\renewcommand{\arraystretch}{1.7}
\textbf{Table 3.} The group classification of class~\eqref{class_f=g}\\[1ex]
\begin{tabular}{|l|c|c|l|}
\hline
N&$f(x)$&$h(x)$&Basis of $A^{\rm max}$\\
\hline
%0&$\forall$&$\forall$&$\partial_t$\\
%\hline
1&$x^{\alpha}e^{\frac{\mu}2x^2}$&$\delta x^{\alpha}e^{\frac{\mu}2x^2+\omega^1}$&
$\partial_t,\,e^{-2\mu t}\bigl[\partial_t-
\mu x\partial_x+\mu\left(2+x\omega^1_x\right)\bigr]\partial_u$\\
\hline
$2.1$&$x^{\alpha}$&$\delta x^{{\alpha}+\frac {\beta}{1-\alpha}}$&$\partial_t,\,2t\partial_t+
x\partial_x-\left(2+\frac {\beta}{1-\alpha}\right)\partial_u$\\
\hline
$2.2$&$x$&$\delta x^{1-\frac {\beta}2\ln x}$&$\partial_t,\,2t\partial_t+
x\partial_x-(2-\beta\ln x)\partial_u$\\
\hline
$3$&$e^{\frac{\mu}2x^2}$&$\delta e^{\frac{\mu}2x^2+\omega^3}$&$\partial_t,\,e^{-\mu t}
\partial_x-e^{-\mu t}\omega^3_x\partial_u$\\
\hline
$4.1$&$e^x$&$\delta e^{\,\rho x}$&$\partial_t,\,
\partial_x+(1-\rho)\partial_u$\\
\hline
$4.2$&$1$&$\delta e^{-\frac {\gamma}2 x^2}$&$\partial_t,\,
\partial_x+\gamma x\partial_u$\\
\hline
5&$e^{\frac{\mu}2x^2}$&$\delta e^{\frac{\mu}2x^2+\omega^5}$&$\partial_t,\,e^{-\mu t}
\partial_x-e^{-\mu t}\omega^5_x\partial_u,$\\
&&&$e^{-2\mu t}\bigl[\partial_t-
\mu x\partial_x+\mu\left(2+x\omega^5_x\right)\bigr]\partial_u$\\
\hline
$6.1$&$e^x$&$\delta e^x$&$\partial_t,\,\partial_x,\,2t\partial_t+
(x-t)\partial_x-2\partial_u$\\
\hline
$6.2$&$1$&$\delta $&$\partial_t,\,\partial_x,\,2t\partial_t+
x\partial_x-2\partial_u$\\
\hline
\end{tabular}
\\[1ex]
\parbox{150mm}{Here $\delta=\pm1$,  $\mu=\pm1\bmod \hat G^{\sim}_1$; $\alpha, \beta, \gamma, \rho$ are arbitrary constants,
$\rho\neq1$, $\alpha^2+\beta^2\neq0$.

$\omega^1=-\int x^{-\alpha} e^{-\frac{\mu}2x^2}\int(\beta x^{-2}+2\mu)x^{\alpha}
e^{\frac{\mu}2x^2}dx\, dx$,\quad
$\omega^3=-\gamma\int e^{-\frac{\mu}2x^2}\int
e^{\frac{\mu}2x^2}dx\, dx$,

 $\omega^5=\omega^3|_{\gamma=2\mu}$,\quad $\omega^i_x=\frac{d\omega^i}{dx}$, i=1,3,5.\, In case 2.1 $\alpha\neq1$.
In case 3 $\gamma\neq2\mu$. In case 4.2 $\gamma\ne0$.}
\end{center}

The kernel of the maximal Lie invariance algebras of equations from class~\eqref{class_f=g}
coincides with the one-dimensional algebra $\langle\partial_t\rangle$.

\section{Construction of  exact solutions via reduction method}
In this section we present an example of finding exact solutions of equations from class~\eqref{class_f=g}
via reduction method. This technique is well known and quite 
algorithmic (see, e.g.,~\cite{Ovsiannikov1982,Olver1986}).

As shown in the previous section, equation~\eqref{case3_eq2} with $\gamma\neq-1$
(Case~4.1 of Table 3 with $\rho=-\gamma$ and $\delta=1$)
admits the two-dimensional (commutative) Lie invariance algebra~$\mathfrak g$ generated by the operators
\[
X_1=\partial_{\tilde t},\quad X_2=\partial_{\tilde x}+(1+\gamma)\partial_{\tilde u}.
\]
A complete list of inequivalent non-zero subalgebras of~$\mathfrak g$ is exhausted by the algebras
$\langle X_1\rangle$, $\langle X_2\rangle$ and $\langle X_1, X_2\rangle$.

Lie reduction of equation~\eqref{case3_eq2} to an algebraic equation
can be made with the two-dimensional subalgebra $\langle X_1, X_2\rangle$
which coincides with the whole algebra~$\mathfrak g$.
The associated ansatz and the reduced algebraic equation have the form

\smallskip

 $\langle X_1,X_2\rangle$:\quad  $\tilde u=(1+\gamma){\tilde x}+C$,\quad
$(1+\gamma)+e^C=0$.

\smallskip

\noindent
The real solution of the reduced equation exists only for $\gamma<-1$.
Substituting the solution $C=\ln|1+\gamma|$ of the reduced algebraic equation into the ansatz,
we construct the exact solution
\begin{equation}\label{sol1_case4}
\tilde u=(1+\gamma){\tilde x}+\ln|1+\gamma|
\end{equation}
of equation~\eqref{case3_eq2} for $\gamma<-1$.

The ansatzes and reduced equations corresponding to the one-dimensional subalgebras from the optimal system
are the following:
\[\begin{array}{l}\langle X_1\rangle:\quad \tilde u=z(y),\quad y={\tilde x};\quad
z_{yy}+z_{y}+e^{-(1+\gamma)y}e^{z}=0;\\[1ex]
\langle X_2\rangle:\quad
\tilde u=(1+\gamma){\tilde x}+z(y),\quad
y=\tilde t; \quad z_{y}=(1+\gamma)+e^{z}.
\end{array}\]
The solution of the latter reduced equation is
$z=\ln\left|\dfrac{\pm(1+\gamma)}{e^{-(y+c)(1+\gamma)}\mp1}\right|$,
where $c$ is an arbitrary constant. Then
\begin{equation}\label{sol2_case4}
\tilde u=(1+\gamma)\tilde x+\ln\left|\dfrac{\pm(1+\gamma)}{e^{-(\tilde t+c)(1+\gamma)}\mp1}\right|
\end{equation} is the corresponding solution of equation~\eqref{case3_eq2}.

Applying the equivalence transformation adduced in Example~1 to~\eqref{sol1_case4} and~\eqref{sol2_case4}
  exact solutions of equation~\eqref{case3_eq1} with complicated coefficients can be easily constructed.

\section{Conclusion}
The complete solution of the group classification problem
for class~\eqref{eqRDfghExp_n=0} became possible only due to using of
the method based on simultaneous application of a mapping between classes
and equivalence transformations.
This method can be applied for solving of similar problems for other classes of differential equations and
extended, e.g., to investigations of 
reduction operators (nonclassical symmetries), 
conservation laws and potential symmetries. 
The usage of transformations from the generalized 
extended equivalence group allows us to present the 
final result in concise form.

\subsection*{Acknowledgments}
The author thanks the
Organizing Committee of the 5th Mathematical Physics
Meeting and especially Prof. Branko Dragovich for hospitality and giving an opportunity to give a talk.
Her participation in the conference was partially supported by CEI and ICTP.
The author is also grateful to Prof. Roman Popovych for useful discussions.

\end{document}